\documentclass[11pt]{article}

\usepackage{arxiv}

\usepackage{amsfonts}
\usepackage{amsmath}
\usepackage{hyperref}

\usepackage{graphicx}

\usepackage{lathalesians-commons}
\usepackage{lathalesians-font}
\usepackage{lathalesians-theorems}

\usepackage{lathalesians-analysis}
\usepackage{lathalesians-linear}
\usepackage{lathalesians-probability}

\usepackage[numbers]{natbib}

\title{Implementing portfolio risk management and hedging in practice}
\author{Paul Alexander Bilokon \\ Thalesians Ltd \\ Level39, One Canada Square \\ Canary Wharf \\ London E14 5AB}
\date{2023.09.27}

\begin{document}

\maketitle

\begin{abstract}
In academic literature portfolio risk management and hedging are often versed in the language of stochastic control and Hamilton--Jacobi--Bellman~(HJB) equations in continuous time. In practice the continuous-time framework of stochastic control may be undesirable for various business reasons. In this work we present a straightforward approach for thinking of cross-asset portfolio risk management and hedging, providing some implementation details, while rarely venturing outside the convex optimisation setting of (approximate) quadratic programming~(QP). We pay particular attention to the correspondence between the economic concepts and their mathematical representations; the abstractions enabling us to handle multiple asset classes and risk models at once; the dimensional analysis of the resulting equations; and the assumptions inherent in our derivations. We demonstrate how to solve the resulting QPs with CVXOPT.
\end{abstract}

\section{Introduction}

In academic literature portfolio risk management and hedging are often (but not always~\cite{Joshi2013, Meucci2005}) versed in the language of stochastic control~\cite{Astrom2006, Yong1999, Pham2009, Kushner2000, Oksendal2019, Oksendal2000} and Hamilton--Jacobi--Bellman~(HJB) equations in continuous time. Indeed, this is the most rigorous and general framework for such considerations. Due to the inherent, natural relationship between stochastic control and reinforcement learning~\cite{Bertsekas2001, Bertsekas2005}, this framework can be relatively easily extended to the numerical methods of reinforcement learning~\cite{Sutton2018, Powell2022} taking care of the more realistic setting involving frictions.

In the author's practical experience of portfolio and risk management on the sell-side and on the buy-side, including at bulge bracket institutions, this approach presents several practical problems. First, some quantitative analysts and quantitative developers~\cite{Derman2007, Joshi2008} come from backgrounds, which exclude stochastic control. For example, many classical computer science degrees do not cover stochastic control (or indeed stochastic analysis) as part of the syllabus. The continuous-time framework is technically complex and requires discretisation at some point anyway. There are relatively few industry-grade solvers for HJB-type problems.

On the other hand, quadratic programming~\cite{Best2010, Cornuejols2018, Gill1982, Boyd2004} is accessible to most majors and requires a relatively limited background in undergraduate linear algebra (rather than graduate-level stochastic analysis and measure theory). It is usually well understood by quantitative analysts and developers irresepective of their academic and professional background. Furthermore, there are fast industry-grade implementations of quadratic program~(QP) solvers, including CPLEX~\cite{IBM2009}, Gurobi~\cite{Gurobi2023}, IMSL~\cite{IMSL1997}, NAG~\cite{NAG2023}, NASOQ~\cite{Cheshmi2020}, OSQP~\cite{OSQP2020}, Xpress~\cite{FICO2023}. Therefore it makes sense to come up with a QP formulation of the portfolio risk management and hedging problem. Even if such approach is imperfect in comparison with the stochastic control approach, it is practical.

Finally, in the medium- to high-frequency trading~(HFT)~\cite{Aldridge2013} applications, the optimiser needs to be very efficient and called sparingly at discrete time intervals. In this setting, the QP formulation is yet again appealing.

In this work we describe the portfolio risk management and hedging framework with sufficient rigour, paying particular attention to the correspondence between the economic concepts and their mathematical representations; the abstractions enabling us to handle multiple asset classes and risk models at once; the dimensional analysis of the resulting equations; and the assumptions inherent in our derivations.

We demonstrate how to solve the resulting QPs with CVXOPT~\cite{Andersen2020}, a free software package for convex optimisation based on the Python programming language. CVXOPT is convenient in the research and development context. In production the reader is advised to use one of the industry-grade C++ or kdb+/q~\cite{Novotny2019} implementations, or a specialised implementation of the solver on an FPGA or ASIC~\cite{Lockwood2012}.

The implementations that we provide here are baseline, pedagogical implementations. In some cases the business requirements may be such that additional frictions may need to be taken into account, in which case the problem ceases to be convex. There are some proprietary tricks that can be applied in these situations. These tricks can significantly impact the profitability and risk profile of a business, but unfortunately they are outside the scope of this work.

\paragraph{Acknowledgements}

We would like to thank Berc Rustem (Department of Computing, Imperial College London) and Martin Zinkin (Qubealgo) for our constructive discussions. The factor abstraction is largely due to Attilio Meucci and follows~\cite{Meucci2005}.

\section{Preliminaries}

In what follows we assume that we are the market maker and the sides of the trades are given from our perspective: buys are when \emph{we} buy (the change in our position has a positive sign), sells are when \emph{we} sell (the change in our position has a negative sign).

We shall denote by $\Nnz$ the set $\{1, 2, 3, \ldots\}$; by $\mathbf{1}_n$ the vector in $\R^n$ whose elements are all ones. The vectors are column vectors by default.

Let us establish conventions for matrix calculus. The \emph{(scalar-by-vector) derivative} of a scalar $y$ with respect to a vector in $\R^n$,
\begin{equation*}
\V{x} =
\left(
  \begin{array}{c}
    x_1 \\
    \vdots \\
    x_n \\
  \end{array}
\right),
\end{equation*}
with $n \in \Nnz$, is written, in \emph{numerator layout notation}, as
\begin{equation*}
\frac{\partial y}{\partial \V{x}} =
\left(
  \begin{array}{ccc}
    \frac{\partial y}{\partial x_1} &\ldots &\frac{\partial y}{\partial x_n}
  \end{array}
\right).
\end{equation*}
Strictly speaking, the result is a matrix in $\R^{1 \times n}$, although we can certainly, and somewhat confusingly, think of it as a vector in $\R^n$. (Confusingly, because of the widespread habit to think of vectors as column vectors by default. To avoid this confusion, we shall favour the matrix view over the vector view whenever there is ambiguity.)

The (vector-by-vector derivative) of a $\R^m$-valued vector function (a vector whose components are functions)
\begin{equation*}
\V{y} =
\left(
  \begin{array}{c}
    y_1 \\
    \vdots \\
    y_m \\
  \end{array}
\right),
\end{equation*}
with $m \in \Nnz$, with respect to an input vector in $\R^n$,
\begin{equation*}
\V{x} =
\left(
  \begin{array}{c}
    x_1 \\
    \vdots \\
    x_n \\
  \end{array}
\right),
\end{equation*}
with $n \in \Nnz$, is written, in numerator layout notation, as
\begin{equation*}
\frac{\partial \V{y}}{\partial \V{x}} =
\left(
  \begin{array}{ccc}
    \frac{\partial y_1}{\partial x_1} & \cdots & \frac{\partial y_1}{\partial x_n} \\
    \vdots & \ddots & \vdots \\
    \frac{\partial y_m}{\partial x_1} & \cdots & \frac{\partial y_m}{\partial x_n} \\
\end{array}
\right).
\end{equation*}
The result is a matrix in $\R^{m \times n}$. It is called the \emph{Jacobian} matrix of $\V{y}$ with respect to $\V{x}$.

We have included these definitions here because they vary across the literature, with some authors preferring the numerator layout notation (as we do throughout this document), while others preferring the denominator layout notation.

Given the numerator layout convention that we have chosen, if $\V{x} \in \R^n$, $\M{A} \in \R^{n \times n}$, then $\frac{\partial}{\partial \V{x}} \transpose{\V{x}} \M{A} \V{x} = \transpose{\V{x}} (\M{A} + \transpose{\M{A}})$. If, moreover, $\M{A}$ is symmetric, then $\frac{\partial}{\partial \V{x}} \transpose{\V{x}} \M{A} \V{x} = 2 \transpose{\V{x}} \M{A}$. If $\V{x} \in \R^n$, $\M{A} \in \R^{m \times n}$, then $\frac{\partial}{\partial \V{x}} \M{A} \V{x} = \M{A}$.

\section{CVXOPT}

CVXOPT's function \textsf{qp} is an interface to \textsf{coneqp} for quadratic programs. It also provides the option of using the quadratic programming solver from MOSEK~\cite{MOSEK2019}.

\begin{verbatim}
cvxopt.solvers.qp(P,q[,G,h,[,A,b[,solver[,initvals]]]])
\end{verbatim}

This solves the convex quadratic program
\begin{equation*}
\begin{aligned}
& \underset{x}{\text{minimise}} & & (1/2)x^T P x + q^T x \\
& \text{subject to} & & Gx \preceq h, \\
&                   & & Ax = b \\
\end{aligned}
\end{equation*}
and its dual.

\section{The portfolio}

Suppose that we are trading $n \in \Nnz$ products whose prices per unit notional\footnote{We shall use the words ``size'' and ``notional'' synonymously.} at time $t$ are given by the vector~$\V{P}_t \in \R^n$. We shall represent the composition of our portfolio, $\pi_t$, in terms of $\V{N}_t \in \R^n$, where the $i$th element, $(\V{N}_t)_i$, is the net notional amount of the $i$th product in the portfolio ($i = 1:n$). Thus the total net notional of our portfolio is given by $N^{\pi}_t = \transpose{\mathbf{1}_n} \V{N}_t \in \R$. We could also consider the weights of the products in our portfolio, $\V{w}_t = \frac{1}{N^{\pi}_t} \V{N}_t \in \R^n$. The value of our portfolio at time $t$ is given by $V_t^{\pi} = \transpose{\V{N}_t} \V{P}_t \in \R$.

This is an appropriate time to comment on the units. The value of the portfolio, $V_t^{\pi}$ is always expressed in units of currency, say, USD or EUR. The units of the prices, $\V{P}_t$, and notionals, $\V{N}_t$, are subject to the conventions of the particular asset class:
\begin{itemize}
\item For equities, $\V{N}_t$ is dimensionless; the notional is expressed as the number of shares. The price, $\V{P}_t$, is in units of currency.
\item For CDS indices, $\V{N}_t$ is expressed in units of currency, as this is the amount on which protection is being bought or sold (e.g. 25,000,000 USD). The price, $\V{P}_t$, on the other hand, is dimensionless, as it is expressed in units of currency per unit notional, which is itself expressed in units of currency.
\end{itemize}

\section{The invariants}

Our market risk consists in our dependence on the prices of the products at time $T + \tau$, where $T$ represents the time when the asset allocation decision is being made and $\tau \in [0, +\infty)$ is our investment horizon.

In order to understand our market risk, we seek to express $\V{P}_t \in \R^n$ in terms of \emph{invariants} --- market variables that exhibit stationary behaviour over time and can be expressed in terms of independent and identically distributed random variables.

Let us denote by $\V{X}_{t, \tau}$ the $\R^n$-valued random vector of invariants at time $t$ for a given investment horizon $\tau$. We express the prices of the products in our portfolio as
\begin{equation*}
\V{P}_t = \V{g}_{\tau}(\V{X}_{t, \tau}),
\end{equation*}
where $\functype[\V{g}_{\tau}]{\R^n}{\R^n}$ is a function that depends on the asset classes of products in our portfolio and our investment horizon. In particular,
\begin{equation*}
\V{P}_{T + \tau} = \V{g}_{\tau}(\V{X}_{T + \tau, \tau}).
\end{equation*}
Notice that we are assuming that, like $\V{P}_t$, $\V{X}_t$ is $n$-dimensional. The reason why this is a sensible assumption will become clear when we consider examples of $\V{g}_{\tau}$ and $\V{X}_{t, \tau}$ for various asset classes.

\section{The factor model}

To get a handle on our market risk, we shall express the vector of invariants as
\begin{equation}
\V{X}_{t, \tau} = \V{h}(\V{F}_{t, \tau}) + \V{U}_{t, \tau},
\label{eq:factor-model}
\end{equation}
where $F_{t, \tau}$ is an $\R^m$-valued ($m \in \Nnz$) vector of common \emph{risk factors} that are responsible for most of the randomness in the market, $U_{t, \tau}$ is a residual vector of perturbations, and $\functype[\V{h}]{\R^m}{\R^n}$ is a function.

\section{Risk}

What is our \emph{risk exposure}, i.e. the sensitivity of the value of our portfolio to our risk factors? It is given by the $\R^m$-valued
\begin{equation*}
\V{r}_{t, \tau} \defeq \frac{\partial V_t^{\pi}}{\partial \V{F}_{t, \tau}} = \frac{\partial}{\partial \V{F}_{t, \tau}} \transpose{\V{N}_t} \V{P}_t = \transpose{\V{N}_t} \frac{\partial}{\partial \V{F}_{t, \tau}} \V{g}(\V{X}_{t, \tau}) = \transpose{\V{N}_t} \frac{\partial \V{g}}{\partial \V{X}_{t, \tau}}(\V{X}_{t, \tau}) \frac{\partial \V{h}}{\partial \V{F}_{t, \tau}}(\V{F}_{t, \tau}).
\end{equation*}

The dependence of our risk exposure on the notional vector is expressed by the matrix of sensitivities, the $m \times n$ Jacobian,
\begin{equation*}
\M{H}_{t, \tau} = \frac{\partial \V{r}_{t, \tau}}{\partial \V{N}_t} = \frac{\partial}{\partial \V{N}_t} \left( \transpose{\V{N}_t} \frac{\partial \V{g}}{\partial \V{X}_{t, \tau}}(\V{X}_{t, \tau}) \frac{\partial \V{h}}{\partial \V{F}_{t, \tau}}(\V{F}_{t, \tau}) \right) = \frac{\partial \V{g}}{\partial \V{X}_{t, \tau}}(\V{X}_{t, \tau}) \frac{\partial \V{h}}{\partial \V{F}_{t, \tau}}(\V{F}_{t, \tau}).
\end{equation*}

Thus the relationship between $\V{r}_{t, \tau}$ and $\M{H}_{t, \tau}$, which follows from their definitions, is given by
\begin{equation}
\V{r}_{t, \tau} = \transpose{\V{N}_t} \M{H}_{t, \tau}.
\label{eq:relationship-between-r-and-H}
\end{equation}

\section{The variance of the value of the portfolio}
\label{sec:variance-of-value-of-portfolio}

We would like to minimise the variance of the value of the portfolio at time $T + \tau$, which represents our market risk. Let us regard $V_t^{\pi}$ as a function of the risk factors: $V_t^{\pi} = V_t^{\pi}(\V{F}_{t, \tau})$. By Theorem~\ref{thm:variance-of-function-of-random-variable} and Remark~\ref{rem:covariance-of-function-of-random-variable},
\begin{equation*}
\Var[V_t^{\pi}] \approx \transpose{\left(\frac{\partial V_t^{\pi}}{\partial \V{F}_{t, \tau}}\right)} \M{C} \frac{\partial V_t^{\pi}}{\partial \V{F}_{t, \tau}} = \transpose{\V{r}_{t, \tau}} \M{C} \V{r}_{t, \tau},
\end{equation*}
where $\M{C} \defeq \Cov(\V{F}_{t, \tau}) \in \R^{m \times m}$ is the covariance matrix.

This result is approximate. We approximated in two places:
\begin{enumerate}
\item In Theorem~\ref{thm:variance-of-function-of-random-variable}, where we disposed of the remainder term of the Taylor series expansion.
\item When we regarded $V_t^{\pi}$ as a function of $\V{F}_{t, \tau}$ alone, whereas it is also a function of another random variable: $\V{U}_{t, \tau}$. We have effectively approximated (\ref{eq:factor-model}) by
    \begin{equation*}
    \V{X}_{t, \tau} \approx \V{h}(\V{F}_{t, \tau}),
    \end{equation*}
    which is sensible if the variance of $\V{U}_{t, \tau}$ is small, i.e. our risk factors are responsible for most of the variance of $\V{X}_{t, \tau}$. To better account for the variance of $\V{U}_{t, \tau}$, we could have used the bivariate Taylor series expansion, instead of the univariate as in the proof of Theorem~\ref{thm:variance-of-function-of-random-variable}.
\end{enumerate}

\section{Minimising risk}

We would like to minimise the variance of the value of our portfolio at our investment horizon, $\Var[V_{T + \tau}^{\pi}]$. Bearing in mind the caveats mentioned in Section~\ref{sec:variance-of-value-of-portfolio}, this is achieved by minimising
\begin{equation*}
\Var[V_{T + \tau}^{\pi}] \approx \transpose{(\V{r}_{T + \tau, \tau} + \M{H}_{T + \tau, \tau} \V{x})} \M{C}_{T + \tau, \tau} (\V{r}_{T + \tau, \tau} + \M{H}_{T + \tau, \tau} \V{x}),
\end{equation*}
where $\M{C}_{T + \tau, \tau} \defeq \Cov(\V{F}_{T + \tau, \tau})$ and $\V{x} \in \R^n$ is the change in position that will minimise the variance of the value of our portfolio. It is $\V{x}$ that we need to find.

Note that, at the time the hedging decision is made, $\V{r}_{T + \tau, \tau}$, $\M{H}_{T + \tau, \tau}$, and $\M{C}_{T + \tau, \tau}$ are not yet known. Therefore we have to make use of forecasts instead and minimise
\begin{equation*}
\transpose{(\hat{\V{r}}_{T + \tau, \tau} + \hat{\M{H}}_{T + \tau, \tau} \V{x})} \hat{\M{C}}_{T + \tau, \tau} (\hat{\V{r}}_{T + \tau, \tau} + \hat{\M{H}}_{T + \tau, \tau} \V{x}),
\end{equation*}
where $\hat{\M{C}}_{T + \tau, \tau} \defeq \Cov(\hat{\V{F}}_{T + \tau, \tau})$. We shall discuss how these forecasts can be obtained later. For the time being, we shall drop the subscripts and superscripts to avoid notational clutter and solve the unconstrained quadratic program~(\textbf{QP})
\begin{equation}
\underset{\V{x}}{\text{minimise}} \, \, \transpose{(\V{r} + \M{H} \V{x})} \M{C} (\V{r} + \M{H} \V{x}),
\label{eq:minimise-risk}
\end{equation}
which can be done analytically using straightforward matrix calculus. First, differentiate with respect to $\V{x}$:
\begin{align*}
\frac{\partial}{\partial \V{x}} \transpose{(\V{r} + \M{H} \V{x})} \M{C} (\V{r} + \M{H} \V{x}) &= \frac{\partial}{\partial \V{x}} \left( \transpose{\V{x}} \transpose{\M{H}} \M{C} \M{H} \V{x} + \transpose{\V{r}} \M{C} \M{H} \V{x} + \transpose{\V{x}} \transpose{\M{H}} \M{C} \V{r} + \transpose{\V{r}} \M{C} \V{r} \right) \\
&= \frac{\partial}{\partial \V{x}} \left( \transpose{\V{x}} \transpose{\M{H}} \M{C} \M{H} \V{x} + 2\transpose{\V{r}} \M{C} \M{H} \V{x} + \transpose{\V{r}} \M{C} \V{r} \right) \\
&= 2 \transpose{\V{x}} \transpose{\M{H}} \M{C} \M{H} + 2 \transpose{\V{r}} \M{C} \M{H}.
\end{align*}
We set this partial derivative to zero and solve for $\V{x}$:
\begin{equation*}
\transpose{\V{x}} \transpose{\M{H}} \M{C} \M{H} = -\transpose{\V{r}} \M{C} \M{H},
\end{equation*}
hence
\begin{equation*}
\transpose{\V{x}} = -\transpose{\V{r}} \M{C} \M{H} \left( \transpose{\M{H}} \M{C} \M{H} \right)^{-1}
\end{equation*}
and
\begin{equation*}
\V{x} = -\left(\transpose{\M{H}} \M{C} \M{H}\right)^{-1} \transpose{\M{H}} \M{C} \V{r} \in \R^n.
\end{equation*}

\section{Minimising risk and symmetric costs}
\label{sec:minimising-risk-and-symmetric-costs}

Let us now incorporate the costs into the optimisation. We now assume that it will cost us $\transpose{\V{c}} \V{x}$ to execute the hedge (change our position by $\V{x}$), where $\V{c} \in \R^n$ are the costs per unit notional. Notice that the costs are symmetric --- they are the same linear factors of $\V{x}$ irrespective of whether we are buying or selling. This assumption may not be realistic in practice.

The problem remains a constrained \textbf{QP}:
\begin{equation*}
\underset{\V{x}}{\text{minimise}} \, \, \transpose{(\V{r} + \M{H} \V{x})} \M{C} (\V{r} + \M{H} \V{x}) + \lambda_c \transpose{\V{c}} |\V{x}|,
\end{equation*}
where $|\cdot|$ denotes the elementwise absolute value.\footnote{At first sight, the problem is an unconstrained \textbf{QP},
\begin{equation*}
\underset{\V{x}}{\text{minimise}} \, \, \transpose{(\V{r} + \M{H} \V{x})} \M{C} (\V{r} + \M{H} \V{x}) + \lambda_c \transpose{\V{c}} \V{x},
\end{equation*}
which can be solved using matrix calculus as before:
\begin{align*}
\frac{\partial}{\partial \V{x}} \left( \transpose{(\V{r} + \M{H} \V{x})} \M{C} (\V{r} + \M{H} \V{x}) + \lambda_c \transpose{\V{c}} \V{x} \right) &= \frac{\partial}{\partial \V{x}} \left( \transpose{\V{x}} \transpose{\M{H}} \M{C} \M{H} \V{x} + \transpose{\V{r}} \M{C} \M{H} \V{x} + \transpose{\V{x}} \transpose{\M{H}} \M{C} \V{r} + \transpose{\V{r}} \M{C} \V{r} + \lambda_c \transpose{\V{c}} \V{x} \right) \\
&= \frac{\partial}{\partial \V{x}} \left( \transpose{\V{x}} \transpose{\M{H}} \M{C} \M{H} \V{x} + 2\transpose{\V{r}} \M{C} \M{H} \V{x} + \transpose{\V{r}} \M{C} \V{r} + \lambda_c \transpose{\V{c}} \V{x} \right) \\
&= 2 \transpose{\V{x}} \transpose{\M{H}} \M{C} \M{H} + 2 \transpose{\V{r}} \M{C} \M{H} + \lambda_c \transpose{\V{c}}.
\end{align*}
Setting this partial derivative to zero and solving for $\V{x}$, we get:
\begin{equation*}
\transpose{\V{x}} \transpose{\M{H}} \M{C} \M{H} = -\left( \transpose{\V{r}} \M{C} \M{H} + \frac{1}{2} \lambda_c \transpose{\V{c}} \right) \left( \transpose{\M{H}} \M{C} \M{H} \right)^{-1},
\end{equation*}
hence
\begin{equation*}
\V{x} = -\left(\transpose{\M{H}} \M{C} \M{H}\right)^{-1} \left( \transpose{\M{H}} \M{C} \V{r} + \frac{1}{2} \lambda_c \V{c} \right) \in \R^n.
\end{equation*}

However, in this form the problem is misspecified: if $\V{c}$ has all positive elements, then we are paying for buying, but \emph{receiving money} for selling!}

Here $\lambda_c \geq 0$ is a nonnegative constant specifying how many units of cash we are prepared to pay for reducing the variance of the value of the portfolio by one unit (in units of value squared).

We can rewrite this problem as
\begin{equation*}
\begin{aligned}
& \underset{\V{x}, \V{v}}{\text{minimise}} & & \transpose{(\V{r} + \M{H}\V{x})} \M{C} (\V{r} + \M{H}\V{x}) + \lambda_c \transpose{(\V{c})} \V{v} + \lambda_0 (\transpose{\V{v}} \V{v} - \transpose{\V{x}} \V{x})\\
& \text{subject to} & & \V{v} \succeq \V{x}, \\
&                   & & \V{v} \succeq -\V{x}.
\end{aligned}
\end{equation*}
The last term was added to regularise the covariance matrix of the overall problem.

Let us write this problem as a standard \textbf{QP}. Setting
\begin{equation*}
\V{x}^a =
\left(
  \begin{array}{c}
    \V{x} \\
    \V{v} \\
  \end{array}
\right),
\end{equation*}
the objective function can be written as
\begin{gather*}
\frac{1}{2} \transpose{(\V{x}^a)}
\left(
  \begin{array}{cc}
     2\transpose{\M{H}} \M{C} \M{H} - \lambda_0 \M{I}_{n \times n} & \M{0}_{n \times n} \\
    \M{0}_{n \times n} & 2 \lambda_0 \M{I}_{n \times n} \\
  \end{array}
\right)
\V{x}^a
+
\transpose{\left(
  \begin{array}{c}
    2 \transpose{\M{H}} \M{C} \V{r} + \lambda_0 \V{c} \\
    \V{0}_{n \times 1}
  \end{array}
\right)}
\V{x}^a
.
\end{gather*}

We can also rewrite the constraints in block matrix form as
\begin{equation*}
\left(
  \begin{array}{ccc}
    \M{I}_{n \times n} & -\M{I}_{n \times n} \\
    -\M{I}_{n \times n} & -\M{I}_{n \times n}
  \end{array}
\right)
\V{x}^a
\preceq
\left(
  \begin{array}{c}
    \V{0}_n \\
    \V{0}_n \\
  \end{array}
\right).
\end{equation*}

\subsection{Using CVXOPT}

Thus we can call
\begin{verbatim}
cvxopt.solvers.qp(P,q[,G,h,[,A,b[,solver[,initvals]]]])
\end{verbatim}
with
\begin{equation*}
\text{\textsf{P}} =
\left(
  \begin{array}{cc}
     2\transpose{\M{H}} \M{C} \M{H} - \lambda_0 \M{I}_{n \times n} & \M{0}_{n \times n} \\
    \M{0}_{n \times n} & 2 \lambda_0 \M{I}_{n \times n} \\
  \end{array}
\right),
\end{equation*}
\begin{equation*}
\text{\textsf{q}} =
\left(
  \begin{array}{c}
    2 \transpose{\M{H}} \M{C} \V{r} + \lambda_0 \V{c} \\
    \V{0}_{n \times 1}
  \end{array}
\right),
\end{equation*}
\begin{equation*}
\text{\textsf{G}} =
\left(
  \begin{array}{ccc}
    \M{I}_{n \times n} & -\M{I}_{n \times n} \\
    -\M{I}_{n \times n} & -\M{I}_{n \times n}
  \end{array}
\right),
\end{equation*}
\begin{equation*}
\text{\textsf{h}} = \left(
  \begin{array}{c}
    \V{0}_n \\
    \V{0}_n
  \end{array}
\right)
\end{equation*}
to find the optimal $\V{x}^a$.

\subsection{When is $\M{P}$ positive definite?}

In this section, and in this section only, $\M{P}$ will denote the parameter of \textsf{cvxopt.solvers.qp} and not the vector of prices.

By Lemma~\ref{lem:eigenvalues-of-diagonal-block-matrix}, the eigenvalues of $\M{P}$ are precisely those of $2\lambda_0 \M{I}_{n \times n}$ and $2\transpose{\M{H}} \M{C} \M{H} - \lambda_0 \M{I}_{n \times n}$, combined. Clearly the eigenvalues of $2\lambda_0 \M{I}_{n \times n}$ are just $2\lambda_0$ repeated $n$ times. To find the eigenvalues of $2 \transpose{\M{H}} \M{C} \M{H} - \lambda_0 \M{I}_{n \times n}$, we solve the characteristic equation
\begin{equation*}
\det((2 \transpose{\M{H}} \M{C} \M{H} - \lambda_0 \M{I}_{n \times n}) - \lambda \M{I}_{n \times n}) = 0
\end{equation*}
for $\lambda$. On observing that
\begin{equation*}
\det((2 \transpose{\M{H}} \M{C} \M{H} - \lambda_0 \M{I}_{n \times n}) - \lambda \M{I}_{n \times n}) = 2^n \det\left(\transpose{\M{H}} \M{C} \M{H} - \left(\frac{1}{2}\lambda_0 + \frac{1}{2}\lambda\right) \M{I}_{n \times n}\right),
\end{equation*}
we notice that this deteminant is zero precisely when
\begin{equation*}
\frac{1}{2}\lambda_0 + \frac{1}{2}\lambda = \lambda'
\end{equation*}
for an eigenvalue, $\lambda'$, of $\transpose{\M{H}} \M{C} \M{H}$. Thus, to summarise, the eigenvalues of $\M{P}$ are:
\begin{itemize}
\item $2\lambda_0$ repeated $n$ times;
\item for $i = 1, \ldots, n$, $2\lambda'_i - \lambda_0$, where $\lambda'_i$ is the $i$th eigenvalue of $\transpose{\M{H}} \M{C} \M{H}$.
\end{itemize}
It is now clear that $\M{P}$ is positive definite iff $0 < \lambda_0 < 2\lambda'_{\text{min}}$, where $\lambda'_{\text{min}}$ is the least eigenvalue of $\transpose{\M{H}} \M{C} \M{H}$ (which is positive because $\transpose{\M{H}} \M{C} \M{H}$ is positive definite by assumption).

\section{Minimising risk and asymmetric costs}
\label{sec:minimising-risk-and-asymmetric-costs}

In Section~\ref{sec:minimising-risk-and-symmetric-costs}, we assumed that the costs would be the same irrespective of the signs of the components of $\V{x}$. In this section we shall develop an approach that will allow us to provide separate costs for buying, $\V{c}^+ \in \R^n$, and for selling, $\V{c}^- \in \R^n$.

To this end we also define $\V{x}^a \in \R^{2n}$ as the block vector
\begin{equation*}
\V{x}^a =
\left(
  \begin{array}{c}
    \V{x}^+ \\
    \V{x}^- \\
  \end{array}
\right),
\end{equation*}
with $\V{x}^+, \V{x}^- \in \R^n$ with all their components nonnegative. $\V{x}^+$ specifies the notional amounts to be bought, $\V{x}^-$ specifies the notional amounts to be bought, for each product.

The optimisation problem now becomes
\begin{equation*}
\begin{aligned}
& \underset{\V{x}^+, \V{x}^-}{\text{minimise}} & & \transpose{(\V{r} + \M{H}(\V{x}^+ - \V{x}^-))} \M{C} (\V{r} + \M{H}(\V{x}^+ - \V{x}^-)) + \lambda_c [\transpose{(\V{c}^+)} \V{x}^+ + \transpose{(\V{c}^-)} \V{x}^-] \\
& \text{subject to} & & \V{x}^+ \succeq 0, \\
&                   & & \V{x}^- \succeq 0.
\end{aligned}
\end{equation*}
Here $\lambda_c \geq 0$ is a nonnegative constant specifying how many units of cash we are prepared to pay for reducing the variance of the value of the portfolio by one unit (in units of value squared).

We note that this is now a \emph{constrained} \textbf{QP}.

There is a problem with this formulation: nothing guarantees that we won't be buying and selling the same product simultaneously, i.e. that, for some $i \in \{1, \ldots, n\}$, $(\V{x}^+)_i > 0$ and $(\V{x}^-)_i > 0$. To address this, we add an additional term, $\lambda_0 \transpose{(\V{x}^+)} \V{x}^-$:
\begin{equation*}
\begin{aligned}
& \underset{\V{x}^+, \V{x}^-}{\text{minimise}} & & \transpose{(\V{r} + \M{H}(\V{x}^+ - \V{x}^-))} \M{C} (\V{r} + \M{H}(\V{x}^+ - \V{x}^-)) + \lambda_c [\transpose{(\V{c}^+)} \V{x}^+ + \transpose{(\V{c}^-)} \V{x}^-] + \lambda_0 \transpose{(\V{x}^+)} \V{x}^- \\
& \text{subject to} & & \V{x}^+ \succeq 0, \\
&                   & & \V{x}^- \succeq 0.
\end{aligned}
\end{equation*}

Let us write this problem as a standard \textbf{QP}. The first term of the objective function can be written as
\begin{align*}
\transpose{(\V{r} + \M{H}(\V{x}^+ - \V{x}^-))} \M{C} (\V{r} + \M{H}(\V{x}^+ - \V{x}^-)) = \transpose{(\M{H}(\V{x}^+ - \V{x}^-))} \M{C} \M{H}(\V{x}^+ - \V{x}^-) + 2 \transpose{\V{r}} \M{C} \M{H}(\V{x}^+ - \V{x}^-) + \transpose{\V{r}} \M{C} \V{r}.
\end{align*}
The last term is a constant and can be dropped from the minimisation. The remaining terms can be rewritten as
\begin{gather*}
\transpose{(\V{x}^+ - \V{x}^-)} \transpose{\M{H}} \M{C} \M{H} (\V{x}^+ - \V{x}^-) + 2 \transpose{\V{r}} \M{C} \M{H}(\V{x}^+ - \V{x}^-) \\
=
\transpose{(\V{x}^a)}
\left(
  \begin{array}{cc}
     \transpose{\M{H}} \M{C} \M{H} & -\transpose{\M{H}} \M{C} \M{H} \\
    -\transpose{\M{H}} \M{C} \M{H} &  \transpose{\M{H}} \M{C} \M{H} \\
  \end{array}
\right)
\V{x}^a
+
\transpose{\left(
  \begin{array}{c}
    2 \transpose{\M{H}} \M{C} \V{r} \\
    -2 \transpose{\M{H}} \M{C} \V{r} \\
  \end{array}
\right)}
\V{x}^a.
\end{gather*}
The second term of the objective function can be written as
\begin{equation*}
\lambda_c [\transpose{(\V{c}^+)} \V{x}^+ + \transpose{(\V{c}^-)} \V{x}^-] = \lambda_c
\transpose{\left(
  \begin{array}{c}
    \V{c}^+ \\
    \V{c}^- \\
  \end{array}
\right)}
\V{x}^a.
\end{equation*}
Finally, the third term of the objective function can be written as
\begin{equation*}
\transpose{(\V{x}^a)}
\left(
  \begin{array}{cc}
    \M{0}_{n \times n} & \lambda_0 \M{I}_{n \times n} \\
    \lambda_0 \M{I}_{n \times n} & \M{0}_{n \times n} \\
  \end{array}
\right)
\V{x}^a.
\end{equation*}
Putting this together, we rewrite the objective function as
\begin{equation*}
\frac{1}{2}
\transpose{(\V{x}^a)}
\left(
  \begin{array}{cc}
     2\transpose{\M{H}} \M{C} \M{H} & -2\transpose{\M{H}} \M{C} \M{H} + 2\lambda_0 \M{I}_{n \times n} \\
    -2\transpose{\M{H}} \M{C} \M{H} + 2\lambda_0 \M{I}_{n \times n} &  2\transpose{\M{H}} \M{C} \M{H} \\
  \end{array}
\right)
\V{x}^a
+
\transpose{\left(
  \begin{array}{c}
    2 \transpose{\V{r}} \M{C} \M{H} + \lambda_c \V{c}^+ \\
    -2 \transpose{\V{r}} \M{C} \M{H} + \lambda_c \V{c}^- \\
  \end{array}
\right)}
\V{x}^a.
\end{equation*}

We can also rewrite the constraints in block matrix form as
\begin{equation*}
\left(
  \begin{array}{ccc}
    -\M{I}_{n \times n} & 0_{n \times n} \\
    0_{n \times n} & -\M{I}_{n \times n}
  \end{array}
\right)
\V{x}^a
\preceq
\left(
  \begin{array}{c}
    \V{0}_n \\
    \V{0}_n \\
  \end{array}
\right).
\end{equation*}

\subsection{Using CVXOPT}

Thus we can call
\begin{verbatim}
cvxopt.solvers.qp(P,q[,G,h,[,A,b[,solver[,initvals]]]])
\end{verbatim}
with
\begin{equation*}
\text{\textsf{P}} =
\left(
  \begin{array}{cc}
     2\transpose{\M{H}} \M{C} \M{H} & -2\transpose{\M{H}} \M{C} \M{H} + 2\lambda_0 \M{I}_{n \times n} \\
    -2\transpose{\M{H}} \M{C} \M{H} + 2\lambda_0 \M{I}_{n \times n} &  2\transpose{\M{H}} \M{C} \M{H} \\
  \end{array}
\right),
\end{equation*}
\begin{equation*}
\text{\textsf{q}} = \left(
  \begin{array}{c}
    2 \transpose{\M{H}} \M{C} \V{r} + \lambda_c \V{c}^+ \\
    -2 \transpose{\M{H}} \M{C} \V{r} + \lambda_c \V{c}^-
  \end{array}
\right),
\end{equation*}
\begin{equation*}
\text{\textsf{G}} = \left(
  \begin{array}{ccc}
    -\M{I}_{n \times n} & 0_{n \times n} \\
    0_{n \times n} & -\M{I}_{n \times n}
  \end{array}
\right),
\end{equation*}
\begin{equation*}
\text{\textsf{h}} = \left(
  \begin{array}{c}
    \V{0}_n \\
    \V{0}_n
  \end{array}
\right)
\end{equation*}
to find the optimal
\begin{equation*}
\V{x} = \left(
  \begin{array}{c}
    \V{x}^+ \\
    \V{x}^-
  \end{array}
\right).
\end{equation*}

\subsection{When is $\M{P}$ positive definite?}

In this section, and in this section only, $\M{P}$ will denote the parameter of \textsf{cvxopt.solvers.qp} and not the vector of prices.

By Lemma~\ref{lem:eigenvalues-of-symmetric-block-matrix}, the eigenvalues of $\M{P}$ are precisely those of $2\lambda_0 \M{I}_{n \times n}$ and $4 \transpose{\M{H}} \M{C} \M{H} - 2\lambda_0 \M{I}_{n \times n}$, combined. Clearly the eigenvalues of $2\lambda_0 \M{I}_{n \times n}$ are just $2\lambda_0$ repeated $n$ times. To find the eigenvalues of $4 \transpose{\M{H}} \M{C} \M{H} - 2\lambda_0 \M{I}_{n \times n}$, we solve the characteristic equation
\begin{equation*}
\det((4 \transpose{\M{H}} \M{C} \M{H} - 2\lambda_0 \M{I}_{n \times n}) - \lambda \M{I}_{n \times n}) = 0
\end{equation*}
for $\lambda$. On observing that
\begin{equation*}
\det((4 \transpose{\M{H}} \M{C} \M{H} - 2\lambda_0 \M{I}_{n \times n}) - \lambda \M{I}_{n \times n}) = 4^n \det\left(\transpose{\M{H}} \M{C} \M{H} - \left(\frac{1}{2}\lambda_0 + \frac{1}{4}\lambda\right) \M{I}_{n \times n}\right),
\end{equation*}
we notice that this deteminant is zero precisely when
\begin{equation*}
\frac{1}{2}\lambda_0 + \frac{1}{4}\lambda = \lambda'
\end{equation*}
for an eigenvalue, $\lambda'$, of $\transpose{\M{H}} \M{C} \M{H}$. Thus, to summarise, the eigenvalues of $\M{P}$ are:
\begin{itemize}
\item $2\lambda_0$ repeated $n$ times;
\item for $i = 1, \ldots, n$, $4\lambda'_i - 2\lambda_0$, where $\lambda'_i$ is the $i$th eigenvalue of $\transpose{\M{H}} \M{C} \M{H}$.
\end{itemize}
It is now clear that $\M{P}$ is positive definite iff $0 < \lambda_0 < 2\lambda'_{\text{min}}$, where $\lambda'_{\text{min}}$ is the least eigenvalue of $\transpose{\M{H}} \M{C} \M{H}$ (which is positive because $\transpose{\M{H}} \M{C} \M{H}$ is positive definite by assumption).

\section{Practical considerations}

The universe of products that we trade may be a proper superset of the universe of products that we use to hedge. This is easily implemented within our framework: compute the risk for the overall portfolio, then restrict the set of products under consideration to the hedging universe. Then the dimension $n$ is the number of products that are used for hedging and the derivations remain valid for this restricted set of products.

\section{Special case: $m=n$, $\M{C}$ and $\M{H}$ diagonal}

Let us now consider the special case when $m = n$ and both $\M{C}$ and $\M{H}$ are diagonal matrices:
\begin{equation*}
\M{C} = \left(
          \begin{array}{ccccc}
            C_1    & 0      & 0      & \cdots & 0      \\
            0      & C_2    & 0      & \cdots & 0      \\
            \vdots & \ddots & \ddots & \ddots & \vdots \\
            0      & 0      & \ddots & \ddots & 0      \\
            0      & 0      & \cdots & 0      & C_n
          \end{array}
        \right),
\quad
\M{H} = \left(
          \begin{array}{ccccc}
            H_1    & 0      & 0      & \cdots & 0      \\
            0      & H_2    & 0      & \cdots & 0      \\
            \vdots & \ddots & \ddots & \ddots & \vdots \\
            0      & 0      & \ddots & \ddots & 0      \\
            0      & 0      & \cdots & 0      & H_n
          \end{array}
        \right).
\end{equation*}
Additionally, we require that $\M{C}$ be positive definite (not just positive semidefinite), so $C_i > 0$ for $i=1,\ldots,n$. We shall also require $H_i > 0$ for $i = 1, \ldots, n$.

In this setting, the unconstrained QP (\ref{eq:minimise-risk}) reduces to the system of scalar optimisation problems without any coupling,
\begin{equation*}
\underset{x_i}{\text{minimise}} \, \, H_i^2 x_i^2 + 2 r_i H_i x_i + r_i^2, \quad i=1,\ldots,n.
\end{equation*}
Using elementary calculus --- solving for $x_i$ the derivative of $H_i^2 x_i^2 + 2 r_i H_i x_i + r_i^2$ with respect to $x$ equated to 0 --- we find the optimal $x_i$: $x_i^* = -r_i / H_i$.

Note that in this case, when the correlations are absent, the sign of $x_i$ (i.e. whether we buy or sell the $i$th product) depends entirely on the signs of $r_i$ and $H_i$: $x_i$ is positive (i.e. we have to buy $|x_i| = x_i$ units of notional of the $i$th product) when $r_i$ and $H_i$ are of different signs; $x_i$ is negative (i.e. we have to sell $|x_i| = -x_i$ units of notional of the $i$th product) when $r_i$ and $H_i$ are of the same sign. (Of course, we are assuming that $r_i \neq 0$, as otherwise there is no risk to hedge.)

For this reason there is no need to augment the costs of buying and selling as we did in Section~\ref{sec:minimising-risk-and-asymmetric-costs}. We can simply set
\begin{equation*}
c_i \defeq
\left\{
  \begin{array}{ll}
    \text{cost of buying 1 unit notional of $i$th product}, & \hbox{$r_i$, $H_i$ of different signs;} \\
    \text{cost of selling 1 unit notional of $i$th product}, & \hbox{$r_i$, $H_i$ of same sign.}
  \end{array}
\right.
\end{equation*}
The objective function then becomes $(r_i + H_i x_i)^2 C_i + \lambda_c c_i x_i$ and the optimisation problem
\begin{equation*}
\underset{x_i}{\text{minimise}} \, \, C_i H_i^2 x_i^2 + (2 C_i r_i H_i + \lambda_c c_i) x_i + C_i r_i^2, \quad i=1,\ldots,n,
\end{equation*}
where $\lambda_c$ is as in Section~\ref{sec:minimising-risk-and-symmetric-costs}.

Again using elementary calculus, we find
\begin{equation*}
x_i^* = -\frac{2C_i r_i H_i + \lambda_c c_i}{2 C_i H_i^2}.
\end{equation*}

\section{Case study: CDS indices, no cross-hedging}

Let us now consider the case of CDS indices. Recall that \emph{credit DV01} (\emph{CDV01}) or \emph{CS01} (\emph{Credit Spread 01}) is defined as the change in price of the CDS contract (of a given notional) for a one basis point increase in spread.

For the European CDS indices (the iTraxx family), we shall take $1,000,000$~EUR as our unit of notional. For the North American CDS indices (the CDX family), we shall take $1,000,000$~USD as our unit of notional. So $N_i$ and $x_i$ will have these units.

The unit of price ($P_i$) will be EUR for iTraxx and USD for CDX. Our invariants ($X_i$) will be credit spreads, whose units are basis points. We shall set $h_i$ in~(\ref{eq:factor-model}) to be the identity map, i.e. our factors will be the same as our invariants, $F_i \equiv X_i$. Since our risk exposure is given by $r_i \defeq \frac{dV^{\pi}}{dF_i}$, its units will be EUR per basis point for iTraxx and USD per basis point for CDX. Since the Jacobian is given by $H_i \defeq \frac{dr_i}{dN_i}$, its units will be $1,000,000^{-1} (\text{basis point})^{-1}$: $H_i$ is the change in price (in EUR for iTraxx, USD for CDX) per 1,000,000 EUR for iTraxx (1,000,000 USD for CDX) notional.

\section{Case study: European government bonds}

Suppose that we have a portfolio of $n$ European government bonds. For $i = 1:n$, the price of the $i$th bond at time $t$ is given by
\begin{equation}
(\V{P}_t)_i = \sum_{j=1}^{l_i} c_{i,j} e^{-(y_t(\tau_{t,i,j}) + \lambda_{t,i}) \tau_{t,i,j}},
\label{eq:bond-pricing}
\end{equation}
where
\begin{itemize}
\item $(\V{P}_t)_i$ is the dirty market price of the $i$th bond;
\item $l_i$ is the number of cashflows for the $i$th bond;
\item $c_{i,j}$ is the $j$th cashflow for the $i$th bond;
\item $\tau_{t,i,j}$ is the duration of the time interval between the time $t$ and the time of the $j$th cashflow of the $i$th bond;
\item $\functype[y_t]{[0, \infty)}{\R}$ is the zero curve at time $t$, which is a function that maps the maturities of the cashflows (the durations of the time intervals between the time $t$ and the times of the cashflows) to the continuously compounded interest rates;
\item $\lambda_{t,i}$ is a parallel vertical shift to the zero curve $y_t$ which is required to equate the right-hand side to the dirty market price of the $i$th bond. We shall refer to $\lambda_{t,i}$ as the \emph{idiosyncratic spread} of the $i$th bond at time $t$.
\end{itemize}

We model the yield curve as
\begin{equation*}
y_t(\tau_{\text{cf}}) = \sum_{k=1}^d \beta_{t,k} f_k(\tau_{\text{cf}}),
\end{equation*}
where $d \in \Nnz$ and, for $k = 1:d$,
\begin{equation*}
\functype[f_k]{[0, \infty)}{\R}
\end{equation*}
are some suitably defined basis functions.

As we are interested in intraday market making, our investment horizon $\tau$ is relatively short, so we can assume that the bond prices are our invariants,
\begin{equation*}
\V{X}_{t,\tau} = \V{P}_t,
\end{equation*}
so $\V{g}_{\tau}$ is the identity map, i.e., for all $\V{x} \in \R^n$, $\V{g}_{\tau}(\V{x}) = \V{x}$, so that
\begin{equation*}
\frac{\partial \V{g}}{\partial \V{X}_{t, \tau}}(\V{X}_{t, \tau}) = \M{I}_n.
\end{equation*}

The risk factors are given by
\begin{equation*}
\M{F}_{t,\tau} =
\left(
  \begin{array}{c}
    \hat{\beta}_{t+T,1} \\
    \vdots \\
    \hat{\beta}_{t+T,d} \\
    \hat{\lambda}_{t+T,1} \\
    \vdots \\
    \hat{\lambda}_{t+T,n} \\
  \end{array}
\right),
\end{equation*}
where
\begin{equation*}
\hat{\beta}_{t+T,1}, \ldots, \hat{\beta}_{t+T,d}, \hat{\lambda}_{t+T,1}, \ldots, \hat{\lambda}_{t+T,n}
\end{equation*}
are, respectively, our forecasts for
\begin{equation*}
\beta_{t+T,1}, \ldots, \beta_{t+T,d}, \lambda_{t+T,1}, \ldots, \lambda_{t+T,n}.
\end{equation*}
Our risk factors explain all of the risk, so
\begin{equation*}
\V{X}_{t, \tau} = \V{h}(\V{F}_{t, \tau}),
\end{equation*}
where $\V{h}$ is given by equation~(\ref{eq:bond-pricing}).

The matrix of sensitivities is given by
\begin{align*}
\M{H}_{t, \tau} &= \frac{\partial \V{g}}{\partial \V{X}_{t, \tau}}(\V{X}_{t, \tau}) \frac{\partial \V{h}}{\partial \V{F}_{t, \tau}}(\V{F}_{t, \tau}) \\
&= \frac{\partial \V{h}}{\partial \V{F}_{t, \tau}}(\V{F}_{t, \tau})
\end{align*}
and the risk exposure can be obtained from~(\ref{eq:relationship-between-r-and-H}).

Thus we need to find
\begin{equation*}
\left( \frac{\partial \V{h}}{\partial \V{F}_{t, \tau}}(\V{F}_{t, \tau}) \right)_{i,j}
\end{equation*}
for $i=1:n$, $j=1:d+n$. Applying the chain rule, for $j=1:d$,
\begin{align*}
\left( \frac{\partial \V{h}}{\partial \V{F}_{t, \tau}}(\V{F}_{t, \tau}) \right)_{i,j} &= \frac{\partial \left( \V{P}_t \right)_i}{\partial \beta_{t, j}}(\V{F}_{t, \tau}) \\
&= -\sum_{k=1}^{l_i} \tau_{t,i,k} c_{i,k} e^{-(y_t(\tau_{t,i,k}) + \lambda_{t,i}) \tau_{t,i,k}} f_j(\tau_{t,i,k}),
\end{align*}
and for $j=d+1:d+n$, setting $j' = j-d$,
\begin{align*}
\left( \frac{\partial \V{h}}{\partial \V{F}_{t, \tau}}(\V{F}_{t, \tau}) \right)_{i,j} &= \frac{\partial \left( \V{P}_t \right)_i}{\partial \lambda_{t, j'}}(\V{F}_{t, \tau}) \\
&=
\left\{
  \begin{array}{ll}
    -\sum_{k=1}^{l_i} \tau_{t,i,k} c_{i,k} e^{-(y_t(\tau_{t,i,k}) + \lambda_{t,i}) \tau_{t,i,k}}, & \hbox{$i=j'$;} \\
    0, & \hbox{otherwise.}
  \end{array}
\right.
\end{align*}

\bibliographystyle{plain}

\begin{thebibliography}{10}

\bibitem{FICO2023}
{\em {FICO} Xpress Optimizer Reference Manual}, 2023.

\bibitem{NAG2023}
{\em The {N}umerical {A}lgorithms {G}roup {NAG} Library Manual, Mark 29.2}, 2023.
\newblock \url{https://support.nag.com/numeric/nl/nagdoc_latest/}.

\bibitem{Aldridge2013}
Irene Aldridge.
\newblock {\em High-Frequency Trading: A Practical Guide to Algorithmic Strategies and Trading Systems}.
\newblock Wiley, 2 edition, 2013.

\bibitem{Andersen2020}
Martin Andersen, Joachim Dahl, and Lieven Vandenberghe.
\newblock {CVXOPT}: Convex optimization.
\newblock {\em Astrophysics Source Code Library}, 2020.

\bibitem{MOSEK2019}
MOSEK ApS.
\newblock {\em The MOSEK optimization toolbox for MATLAB manual. Version 9.0.}, 2019.

\bibitem{Astrom2006}
Karl~J. Astrom.
\newblock {\em Introduction to Stochastic Control Theory}.
\newblock Dover, 2006.

\bibitem{Bertsekas2001}
Dimitri~P. Bertsekas.
\newblock {\em Dynamic programming and optimal control, Volume~{I}}.
\newblock Athena Scientific, Belmont, MA, 2001.

\bibitem{Bertsekas2005}
Dimitri~P. Bertsekas.
\newblock {\em Dynamic programming and optimal control, Volume~{II}}.
\newblock Athena Scientific, Belmont, MA, 2005.

\bibitem{Best2010}
Michael~J. Best.
\newblock {\em Portfolio Optimization}.
\newblock {CRC} Press, 2010.

\bibitem{Boyd2004}
Stephen Boyd and Lieven Vandenberghe.
\newblock {\em Convex Optimization}.
\newblock Cambridge University Press, 2004.

\bibitem{Cheshmi2020}
Kazem Cheshmi, Danny~M. Kaufman, Shoaib Kamil, and Maryam~Mehri Dehnavi.
\newblock {NASOQ}.
\newblock {\em {ACM} Transactions on Graphics}, 39(4), aug 2020.

\bibitem{Cornuejols2018}
Gerard Cornuejols, Javier Pena, and Reha Tutuncu.
\newblock {\em Optimization Methods in Finance}.
\newblock Cambridge University Press, 2 edition, 2018.

\bibitem{Derman2007}
Emanuel Derman.
\newblock {\em My Life as a Quant: Reflections on Physics and Finance}.
\newblock Wiley, 2007.

\bibitem{Gill1982}
Philip~E. Gill, Walter Murray, and Margaret~H. Wright.
\newblock {\em Practical Optimization}.
\newblock Emerald Group Publishing Limited, 1982.

\bibitem{Gurobi2023}
{Gurobi Optimization, LLC}.
\newblock {Gurobi Optimizer Reference Manual}, 2023.

\bibitem{IBM2009}
{IBM}.
\newblock {V12.1}: User's manual for {CPLEX}.
\newblock Technical report, {IBM}, 2009.

\bibitem{IMSL1997}
IMSL.
\newblock {\em {IMSL} {STAT/LIBRARY}}.
\newblock {V}isual {N}umerics {I}nc., Houston, Texas, {USA}, 1997.
\newblock \url{http://www.vni.com/books/dod/pdf/STATVol_2.pdf}.

\bibitem{Joshi2008}
Mark Joshi.
\newblock On becoming a quant.
\newblock \url{http://www.maths.usyd.edu.au/u/UG/SM/MATH3075/r/Joshi_2008.pdf}, 2008.

\bibitem{Joshi2013}
Mark~S. Joshi and Jane~M. Paterson.
\newblock {\em Introduction to Mathematical Portfolio Theory}.
\newblock International Series on Actuarial Science. Cambridge University Press, 2013.

\bibitem{Kushner2000}
Harold~J. Kushner and Paul Dupuis.
\newblock {\em Numerical Methods for Stochastic Control Problems in Continuous Time}.
\newblock Springer, 2 edition, 2000.

\bibitem{Lockwood2012}
John~W. Lockwood, Adwait Gupte, Nishit Mehta, Michaela Blott, Tom English, and Kees Vissers.
\newblock A low-latency library in {FPGA} hardware for high-frequency trading ({HFT}).
\newblock In {\em {IEEE} 20th Annual Symposium on High-Performance Interconnects}, pages 9--16, 2012.

\bibitem{Meucci2005}
Attilio Meucci.
\newblock {\em Risk and Asset Allocation}.
\newblock Springer Finance. Springer, 2005.

\bibitem{Novotny2019}
Jan Novotny, Paul~Alexander Bilokon, Aris Galiotos, and Fr{\'e}d{\'e}ric D{\'e}l{\`e}ze.
\newblock {\em Machine Learning and Bid Data with {kdb+/q}}.
\newblock Wiley, 2019.

\bibitem{Oksendal2000}
Bernt {\O}ksendal.
\newblock {\em Stochastic Differential Equations: An Introduction with Applications}.
\newblock Universitext. Springer, 6 edition, 2000.

\bibitem{Oksendal2019}
Bernt {\O}ksendal and Agnes Sulem.
\newblock {\em Applied Stochastic Control of Jump Diffusions}.
\newblock Springer, 3 edition, 2019.

\bibitem{Pham2009}
Huyen Pham.
\newblock {\em Continuous-time Stochastic Control and Optimization with Financial Applications}.
\newblock Springer, 2009.

\bibitem{Powell2022}
Warren~B. Powell.
\newblock {\em Reinforcement Learning and Stochastic Optimization: A Unified Framework for Sequential Decisions}.
\newblock Wiley, 2022.

\bibitem{OSQP2020}
B.~Stellato, G.~Banjac, P.~Goulart, A.~Bemporad, and S.~Boyd.
\newblock {OSQP}: an operator splitting solver for quadratic programs.
\newblock {\em Mathematical Programming Computation}, 12(4):637--672, 2020.

\bibitem{Sutton2018}
Richard~S. Sutton and Andrew~G. Barto.
\newblock {\em Reinforcement Learning: An Introduction}.
\newblock MIT Press, 2 edition, 2018.

\bibitem{Yong1999}
Jiongmin Yong and Xun~Yu Zhou.
\newblock {\em Stochastic Controls: {H}amiltonian Systems and {HJB} Equations}.
\newblock Springer, 1999.

\end{thebibliography}

\appendix

\section{Auxiliary results}

\begin{theorem}[The variance of a function of a random variable]
Let $X$ be a real-valued random variable with known finite expected value and finite nonzero variance and let $\functype[f]{\R}{\R}$ be an integrable function. Then
\begin{equation*}
\Var[f(X)] \approx f'(\E[X])^2 \Var[X].
\end{equation*}
\begin{proof}
The following proof is due to Tomek Tarczynski.\footnote{See Tomek's post \atitle{Variance of a function of one random variable} on \atitle{CrossValidated}: \url{http://stats.stackexchange.com/questions/5782/variance-of-a-function-of-one-random-variable}}

By Chebyshev's inequality for random variables with finite variance, for any real $c > 0$,
\begin{equation*}
\Prob[|X - \E[X]| > c] \leq \frac{1}{c} \Var[X],
\end{equation*}
so for any $\epsilon > 0$ we can find a large enough $c$ so that
\begin{equation*}
\Prob[X \in [\E[X] - c, \E[X] + c]] = \Prob[|X - \E[X]| \leq c] < 1 - \epsilon.
\end{equation*}

Let us estimate $\E[f(X)]$. We can write it as
\begin{equation}
\E[f(X)] = \int_{|x - \E[X]| \leq c} f(x) \, dF(x) + \int_{|x - \E[X]| > c} f(x) \, dF(x),
\label{eq:expected-value-of-f-in-terms-of-integrals}
\end{equation}
where $F$ is the distribution function of $X$.

Since the domain of the first integral is the bounded closed interval $[\E[X] - c, \E[X] + c]$, we can apply the Taylor series expansion:
\begin{equation*}
f(x) = f(\E[X]) + f'(\E[X])(x - \E[X]) + \frac{1}{2} f''(\E[X])(x - \E[X])^2 + \frac{1}{3!} f'''(\alpha)(x - \E[X])^3,
\end{equation*}
where $\alpha \in [\E[X] - c, \E[X] + c]$, and the equality holds for all $x \in [\E[X] - c, \E[X] + c]$. Here we took only four terms in the Taylor series expansion, but in general we can take as many as needed, as long as the function $f$ is smooth enough.

Substituting this formula into~(\ref{eq:expected-value-of-f-in-terms-of-integrals}), we get
\begin{gather*}
\E[f(X)] = \int_{|x - \E[X]| \leq c} f(\E[X]) + f'(\E[X])(x - \E[X]) + \frac{1}{2} f''(\E[X])(x - \E[X])^2 \, dF(x) \\
+ \int_{|x - \E[X]| \leq c} \frac{1}{3!} f'''(\alpha) (x - \E[X])^3 \, dF(x) \\
+ \int_{|x - \E[X]| > c} f(x) \, dF(x).
\end{gather*}

Increasing the domain of integration, we obtain
\begin{equation}
\E[f(X)] = f(\E[X]) + \frac{1}{2} f''(\E[X])\E[X - \E[X]]^2 + R_3
\label{eq:approximate-for-expected-value-of-f}
\end{equation}
where
\begin{gather*}
R_3 = \frac{1}{3!} f'''(\alpha) \E[(X - \E[X])^2] \\
+ \int_{|x - \E[X]| > c} \left( f(\E[X]) + f'(\E[X])(x - \E[X]) + \frac{1}{2} f''(\E[X])(x - \E[X])^2 + f(X) \right) \, dF(x).
\end{gather*}

Under some moment conditions, we can show that the second term of this remainder is as large as $\mathbb{P}[|X - \E[X]| > c]$, which is generally small. The first term remains, therefore the quality of the approximation depends on $\E[(X - \E[X])^3]$ and the behaviour of the third derivative of $f$ on bounded intervals. Such approximation should work particularly well for random variables with zero third central moment, such as the normal distribution.

To obtain an approximation for the variance of $f(X)$, we subtract (\ref{eq:approximate-for-expected-value-of-f}) from the Taylor series expansion for $f(x)$ and square the difference:
\begin{equation*}
\Var[f(X)] = \E[(f(X) - \E[f(X)])^2] = (f'(\E[X]))^2 \Var[X] + T_3,
\end{equation*}
where $T_3$ involves the central moments $\E[(X - \E[X])^k]$ for $k \geq 4$.
\end{proof}
\label{thm:variance-of-function-of-random-variable}
\end{theorem}

\begin{remark}
Theorem~\ref{thm:variance-of-function-of-random-variable} generalises to the $\R^n$-valued random variable $\V{X}$, $n \in \Nnz$, and $\functype[\V{f}]{\R^n}{\R^n}$:
\begin{equation*}
\Var[\V{f}(X)] \approx \transpose{\left(\frac{\partial \V{f}}{\partial \V{X}} (\E[\V{X}])\right)} \Var[\V{X}] \frac{\partial \V{f}}{\partial \V{X}} (\E[\V{X}]).
\end{equation*}
\label{rem:covariance-of-function-of-random-variable}
\end{remark}

\begin{lemma}
Let $\M{M}$ be the block matrix
\begin{equation*}
\M{M} =
\left(
  \begin{array}{cc}
    \M{A} & \M{0}_{n \times m} \\
    \M{0}_{m \times n} & \M{B} \\
  \end{array}
\right)
\end{equation*}
with $\M{A} \in \R^{n \times n}$, $\M{B} \in \R^{m \times m}$. The eigenvalues of $\M{M}$ are precisely those of the matrices $\M{A}$ ($n$ eigenvalues, some of the possibly repeated) and $\M{B}$ ($m$ eigenvalues, some of them possibly repeated).
\begin{proof}
It is well known that for all square matrices $\M{A}$ and $\M{B}$ of equal dimensions, the following holds:
\begin{equation*}
\det
\left(
  \begin{array}{cc}
    \M{A} & \M{0}_{n \times m} \\
    \M{0}_{m \times n} & \M{B} \\
  \end{array}
\right)
= \det(\M{A}) \det(\M{B}).
\end{equation*}
Therefore, the characteristic polynomial of this block matrix is given by
\begin{equation*}
\det \left(
       \begin{array}{cc}
         \M{A} - \lambda \M{I}_{n \times n} & \M{0}_{n \times m} \\
         \M{0}_{m \times n} & \M{B} - \lambda \M{I}_{m \times m} \\
       \end{array}
     \right)
= \det(\M{A} - \lambda \M{I}_{n \times n}) \det(\M{A} - \lambda \M{I}_{m \times m}).
\end{equation*}
It follows that the eigenvalues $\M{M}$ are precisely those of the matrices $\M{A}$ and $\M{B}$, combined.
\end{proof}
\label{lem:eigenvalues-of-diagonal-block-matrix}
\end{lemma}

\begin{lemma}
Let $\M{M}$ be the block matrix
\begin{equation*}
\M{M} =
\left(
  \begin{array}{cc}
    \M{A} & \M{B} \\
    \M{B} & \M{A} \\
  \end{array}
\right)
\end{equation*}
with $\M{A}, \M{B} \in \R^n$, $n \in \Nnz$. The eigenvalues of $\M{M}$ are precisely those of the matrices $\M{A} + \M{B}$ ($n$ eigenvalues, some of them possibly repeated) and $\M{A} - \M{B}$ ($n$ eigenvalues, some of them possibly repeated).
\begin{proof}
It is well known that for all square matrices $\M{A}$ and $\M{B}$ of equal dimensions, the following holds:
\begin{equation*}
\det \left(
       \begin{array}{cc}
         \M{A} & \M{B} \\
         \M{B} & \M{A} \\
       \end{array}
     \right)
= \det(\M{A} - \M{B}) \det(\M{A} + \M{B}).
\end{equation*}
Therefore, the characteristic polynomial of this block matrix is given by
\begin{equation*}
\det \left(
       \begin{array}{cc}
         \M{A} - \lambda \M{I}_{n \times n} & \M{B} \\
         \M{B} & \M{A} - \lambda \M{I}_{n \times n} \\
       \end{array}
     \right)
= \det((\M{A} - \M{B}) - \lambda \M{I}_{n \times n}) \det((\M{A} + \M{B}) - \lambda \M{I}_{n \times n}).
\end{equation*}
It follows that the eigenvalues $\M{M}$ are precisely those of the matrices $\M{A} + \M{B}$ and $\M{A} - \M{B}$, combined.
\end{proof}
\label{lem:eigenvalues-of-symmetric-block-matrix}
\end{lemma}

\end{document}